\documentclass[aps,prd,preprint,showpacs,nofootinbib,eqsecnum,amsmath,amssymb]{revtex4-1}
\usepackage[colorlinks=true,urlcolor=blue,linkcolor=blue,citecolor=blue]{hyperref}
\usepackage{amsmath}
\usepackage{amssymb}
\usepackage{epsf}
\usepackage{color}
\usepackage{dcolumn}
\usepackage{bm}
\newcommand{\aT}{\nearrow\!\!\!\!\!\!\!T}

\begin{document}

\title{Quantum hydrodynamics of spinning particles in electromagnetic and torsion fields}

\author{\firstname{Mariya Iv.}~\surname{Trukhanova}}
\email{trukhanova@physics.msu.ru}
\affiliation{Faculty of Physics, M. V. Lomonosov Moscow State University,
Leninskie Gory, 119991 Moscow, Russia}
\affiliation{Theoretical Physics Laboratory, Nuclear Safety Institute,
Russian Academy of Sciences, B. Tulskaya 52, 115191 Moscow, Russia}

\author{\firstname{Yuri N.}~\surname{Obukhov}}
\email{obukhov@ibrae.ac.ru}
\affiliation{Theoretical Physics Laboratory, Nuclear Safety Institute,
Russian Academy of Sciences, B. Tulskaya 52, 115191 Moscow, Russia}


\begin{abstract}
We develop a many-particle quantum-hydrodynamical model of fermion matter interacting with the external classical electromagnetic and gravitational/inertial and torsion fields. The consistent hydrodynamical formulation is constructed for the many-particle quantum system of Dirac fermions on the basis of the nonrelativistic Pauli-like equation obtained via the Foldy-Wouthuysen transformation. With the help of the Madelung decomposition approach, the explicit relations between the microscopic and macroscopic fluid variables are derived. The closed system of equations of quantum hydrodynamics encompasses the continuity equation, and the dynamical equations of the momentum balance and the spin density evolution. The possible experimental manifestations of the torsion in the dynamics of spin waves is discussed.
\end{abstract}

\maketitle

\section{Introduction}

Spin (an intrinsic angular momentum) is an important physical property of matter, associated with rotation, which is considered as the source of the gravitational field \cite{1,2,21,3,4,41,5,6,7,71} in the framework of the gauge gravity approach. The spin angular momentum, together with the energy-momentum current, determines the geometrical structure of the spacetime manifold and predicts nontrivial post-Riemannian deviations from Einstein's general relativity (GR) theory. The development of the gravitational theory with torsion has a long history, going back to 1922 when \'Elie Cartan \cite{1} introduced the corresponding geometrical formalism. Since the 1960s, the interest in the theory of gravitation with spin and torsion based on the Riemann-Cartan geometry had considerably grown, and the consistent formalism was developed \cite{2,21,3,4,41,5}. A further generalization of the gauge gravitational theory takes into account the additional microstructural physical properties of matter (encompassing the intrinsic shear and the dilation currents along with the spin) as the sources of gravity, that results in the extension of the spacetime structure to the metric-affine geometry \cite{6}. An exhaustive overview of historic developments can be found in \cite{7,71}. 

The study of dynamics of the spinning matter on the Riemann-Cartan spacetime represents a nontrivial problem which is of interest both theoretically and experimentally. Quoting Einstein \cite{Einstein}, ``... the question whether this continuum has a Euclidean, Riemannian, or any other structure is a question of physics proper which must be answered by experience, and not a question of a convention to be chosen on grounds of mere expediency.'' It is now well established that the spacetime torsion can only be detected with the help of the spin \cite{YS,HOP,OP}. The early theoretical analysis of the possible experimental manifestations of the torsion field at low energies can be found in \cite{9}.

By noticing that the spin and the energy-momentum tensors are the two Noether currents for the Poincar\'e group, one can develop a natural formulation of the theory of gravity with torsion as a local gauge theory for the Poincar\'e spacetime symmetry \cite{yno1,yno2,yno3}. This underlies the previous study of the quantum dynamics of a Dirac particle in the Poincar\'e gauge gravitational field \cite{8,GOS}, where both the minimal and nonminimal coupling of the Dirac fermion with the electromagnetic and the gauge gravitational fields was comprehensively analysed for an arbitrary spacetime geometry with the curvature and torsion. It was demonstrated that the Pauli-like equation for the spinning particles contains new torsion-dependent terms which could give rise to the physical effects competing with the electromagnetic ones.

The study of the spin-torsion coupling obviously fits into the general context of the theoretical and experimental research of spin-dependent long-range forces \cite{93,94,95,96,10,11,12,13,GPS,14,Mohanty,Belyaev,VAK,VAK2,Heckel,Kimball,Ivanov1,Ivanov2,Fabbri,15}. Certain extensions of the Standard Model in the high energy particle physics predict the existence of new particles, in particular, of the light pseudoscalar bosons (such as goldstones, axions \cite{Moody,axion1,axion2}, arions \cite{151,152}, etc) that may give rise to the spin-spin interactions of various kinds. An exchange via such light bosons between two fermions is qualitatively described by a magnetic dipole-dipole type potential. Different methods were proposed to detect these spin-spin interactions, including ferromagnetic detectors with a highly sensitive two-channel UHF receiver \cite{152}, paramagnetic salt with a dc SQUID used in a rotating copper mass \cite{94,95}, examining the hyperfine resonances for $^9$Be$^+$ ions stored in the Penning ion trap \cite{96}, and even treating the Earth as a polarized spin source \cite{10}. In the recent experiment \cite{11}, the transversely polarized slow neutrons were used in an attempt to observe a possible spin rotation of neutrons that traversed a meter of liquid $^4$He under the action of the torsion field. Ultracold neutrons provide a convenient tool, with the quantum gravitational states of ultracold neutrons being sensitive to the post-Riemannian contributions \cite{Ivanov2}. On the theoretical side, the covariant multipolar technique was used for the analysis of the equations of motion of test bodies with spin for a very general class of gravitational theories with the minimal and nonminimal coupling \cite{12,13}. An interesting practical realization of theoretical findings has been recently proposed as a new Gravity Probe Spin space mission using mm-scale ferromagnetic gyroscopes in orbit around the Earth \cite{GPS}. Typically, the predicted spin-torsion effects are expected to be quite small and difficult to observe experimentally, however, one can set the experimental bounds on the spin-torsion coupling constants and on the torsion field as well \cite{14,Mohanty,Belyaev,VAK,VAK2,Heckel,Kimball,Ivanov1,Ivanov2,Fabbri,15}.

Here we for the first time develop the quantum hydrodynamics for the many-particle system of massive Dirac fermion spin-1/2 particles interacting with external electromagnetic, metric gravitational/inertial and torsion fields on the basis of the earlier analysis \cite{8}.

This article is organized as follows. In Sec.~\ref{Sec2} we formulate a Pauli-type one-particle equation for a Dirac fermion moving on the background of gravitational and electromagnetic fields. In Sec.~\ref{Sec3} we introduce the many-particle Pauli-like equation and construct the many-particle quantum hydrodynamics for the non-relativistic particles in the external classical fields. We derive the system of hydrodynamical equations and analyze the structure of force fields in these equations. In Sec.~\ref{Sec4} we apply Madelung's decomposition for the spinor wave function to get the basic physical quantities in macroscopic form. Finally, in Sec. \ref{Sec5} we discuss possible experimental manifestations of the results obtained in this article, and Sec. \ref{Sec6} contains our conclusions.

Our basic conventions and notations are the same as in Ref. \cite{6}. In particular, the world indices are labeled by Latin letters $i,j,k,\ldots = 0,1,2,3$ (for example, the local spacetime coordinates $x^i$ and the holonomic coframe $dx^i$), whereas we reserve Greek letters for tetrad indices, $\alpha,\beta,\ldots = 0,1,2,3$ (e.g., the anholonomic coframe $\vartheta^\alpha = e^\alpha_idx^i$). In order to distinguish separate tetrad indices we put hats over them. Finally, spatial indices are denoted by Latin letters from the beginning of the alphabet, $a,b,c,\ldots = 1,2,3$. The metric of the Minkowski spacetime reads $g_{\alpha\beta} = {\rm diag}(c^2, -1, -1, -1)$, and the totally antisymmetric Levi-Civita tensor $\eta_{\alpha\beta\mu\nu}$ has the only nontrivial component $\eta_{\hat{0}\hat{1}\hat{2}\hat{3}} = c$, so that $\eta_{\hat{0}abc} = c\varepsilon_{abc}$ with the three-dimensional Levi-Civita tensor $\varepsilon_{abc}$. The spatial components of the tensor objects are raised and lowered with the help of the Euclidean 3-dimensional metric $\delta_{ab}$.

\section{Pauli equation for the system with spin-torsion coupling}\label{Sec2}

\subsection{Poincar\'e gauge gravity theory: the basics}

Recalling that the Standard Model in the fundamental particle physics is formulated as a gauge theory for the internal unitary symmetry groups, one may say that, apart from the gravitational interaction, the gauge-theoretic approach underlies the modern physics. There exist, however, a natural extension of Einstein's GR that is based on the Poincar\'e symmetry group $G = T_4 \rtimes SO(1,3)$, the semi-direct product of the four-parameter translation group $T_4$ and the six-parameter Lorentz group $SO(1,3)$, with the energy-momentum current and the spin angular momentum current as the sources of the gravitational field \cite{7,71,yno1,yno2,yno3}.

The gauge fields act as mediators of physical interactions for the fermion matter source. Specializing to the electromagnetic and gravitational interactions, in the framework of the standard Yang-Mills-Sciama-Kibble approach \cite{71}, one then describes electromagnetism by the $U(1)$ gauge field potential $A_i$, and in similar way, one describes gravity by the Poincar\'e gauge potentials $e^\alpha_i$ and $\Gamma_i{}^{\alpha\beta}$. Geometrically, the 4 potentials $e^\alpha_i$ of the translation subgroup $T_4$ are naturally interpreted as the coframe (or the tetrad) field of a physical observer on the spacetime manifold $M_4$, whereas the 6 potentials $\Gamma_i{}^{\alpha\beta} = -\,\Gamma_i{}^{\beta\alpha}$ for the Lorentz subgroup $SO(1,3)$ are identified with the local connection that introduces the parallel transport on the spacetime $M_4$.

The multiplet of gauge potentials,
\begin{equation}
\left\{\,A_i,\quad e_i^\alpha,\quad \Gamma_i{}^{\alpha\beta}\,\right\},\label{pot}
\end{equation}
determines the corresponding multiplet of the ``Yang-Mills'' gauge field strengths:
\begin{eqnarray}
F_{ij} &=& \partial_iA_j - \partial_jA_i,\label{Max}\\
T_{ij}{}^\alpha &=& \partial_ie^\alpha_j - \partial_je^\alpha_i
+ \Gamma_{i\beta}{}^\alpha e^\beta_j - \Gamma_{j\beta}{}^\alpha e^\beta_i,\label{Tor}\\
R_{ij}{}^{\alpha\beta} &=& \partial_i\Gamma_j{}^{\alpha\beta}- \partial_j\Gamma_i{}^{\alpha\beta}
+ \Gamma_{i\gamma}{}^\beta\Gamma_j{}^{\alpha\gamma} - \Gamma_{j\gamma}{}^\beta\Gamma_i{}^{\alpha\gamma}.\label{Cur}
\end{eqnarray}
Thereby, we derive the Maxwell tensor $F_{ij}$ as the $U(1)$ gauge field strength for the electromagnetic field, and the spacetime torsion tensor $T_{ij}{}^\alpha$ and the curvature tensor $R_{ij}{}^{\alpha\beta} = -\,R_{ij}{}^{\beta\alpha}$ as the two Poincar\'e ($T_4$ ``translational'' and $SO(1,3)$ ``rotational'', respectively) gauge field strengths for the gravitational field.

The nontrivial ``mixed'' form of the torsion (\ref{Tor}) is explained by the semi-direct structure of the Poincar\'e symmetry group. The resulting Riemann-Cartan geometry on the spacetime $M_4$ is characterized by the nonvanishing torsion and curvature, whereas in the special case $T_{ij}{}^\alpha = 0$ we recover the Riemannian geometry, and for $R_{ij}{}^{\alpha\beta} = 0$ one finds the Weitzenb\"ock space of distant parallelism.

Here we do not discuss the construction of the complete dynamical scheme of the Poincar\'e gauge theory that requires the introduction of the corresponding gravitational field Lagrangian, and consider the electromagnetic and the gravitational fields as a non-dynamical background. It is important to recall, though, that the variation of the Lagrange density of matter with respect to the gauge field potentials (\ref{pot}) gives rise to the corresponding dynamical currents: the electric current, the canonical energy-momentum tensor, and the spin angular momentum tensor, respectively. Further details can be found in \cite{7,71,yno1,yno2,yno3}, and we conclude this section with the following technical points which are needed for the subsequent discussion.

One can decompose the local Lorentz connection into the sum
\begin{equation}
\Gamma_{i}{}^{\alpha\beta} = \widetilde{\Gamma}_{i}{}^{\alpha\beta} - K_{i}{}^{\alpha\beta}\label{GGK}
\end{equation}
of the Riemannian connection (denoted by the tilde), which is torsionless $\partial_ie^\alpha_j - \partial_je^\alpha_i + \widetilde{\Gamma}_{i\beta}{}^\alpha e^\beta_j - \widetilde{\Gamma}_{j\beta}{}^\alpha e^\beta_i = 0$ and metric-compatible, plus the post-Riemannian contortion tensor,
\begin{equation}
K_{i\alpha\beta} = {\frac 12}\left(T_{\alpha\beta i} - T_{i\alpha\beta}
+ T_{i\beta\alpha}\right).\label{contortion}
\end{equation}

On the other hand, the torsion tensor $T_{\mu\nu}{}^\alpha = e_\mu^ie_\nu^jT_{ij}{}^\alpha$ can be decomposed into the three irreducible parts
\begin{equation}\label{Tdec}
T_{\mu\nu}{}^\alpha = \frac{1}{3}(\delta^\alpha_\mu T_\nu - \delta^\alpha_\nu T_\mu)
+ {\frac 13}\eta_{\mu\nu\lambda}{}^\alpha \check{T}^\lambda + \aT_{\mu\nu}{}^\alpha,
\end{equation}
where $\aT_{\mu\nu}{}^\alpha$ is the trace-free and axial trace-free tensor, the torsion trace vector $T_\mu = T_{\alpha\mu}{}^\alpha$, and the axial trace vector
\begin{equation}\label{Taxial}
\check{T}^\alpha = -\,{\frac 12}\,\eta^{\alpha\mu\nu\lambda}T_{\mu\nu\lambda},
\end{equation}
with the totally antisymmetric Levi-Civita tensor $\eta^{\alpha\mu\nu\lambda}$.

\subsection{Hamiltonian for the Dirac fermion}

The Pauli-like equation for a fermion particle, moving under the action of the torsion field had been derived in \cite{9} for the flat Minkowski spacetime, and in \cite{8} for an arbitrary curved space background. The relativistic dynamics of the Dirac particle with spin $1/2$, electric charge $q$, and mass $m$ minimally coupled to the gravitational and electromagnetic fields is described by the invariant action 
\begin{equation}
S = \int{d^4x \sqrt{-g} L},
\end{equation}
where the Lagrangian of the spinor wave function $\psi$ and $\overline{\psi} = \psi^{\dag}\gamma^{\hat 0}$ has the form 
\begin{equation}\label{Ldir}
L = {\frac{i\hbar}{2}}\left(\overline{\psi}\gamma^\alpha D_\alpha\psi
- D_\alpha\overline{\psi}\gamma^\alpha\psi\right) - mc\,\overline{\psi}\psi\,.
\end{equation}
The spinor covariant derivative describes the minimal coupling of the charged Dirac particle with the external electromagnetic and gravitational gauge fields (\ref{pot})
\begin{equation}
D_\alpha = e_\alpha^i \left(\partial _i - {\frac {iq}{\hbar}}\,A_i
+ {\frac i4}\Gamma_i{}^{\beta\gamma}\sigma_{\beta\gamma}\right).\label{eqin2} 
\end{equation}
Here $c$ and $\hbar$ are the speed of light and Planck's constant, respectively, the 4-potential of the electromagnetic field $A_i = (-\phi, \bm{A})$ encompasses the scalar $\phi$ and vector $\bm{A}$ potentials, and $\sigma_{\alpha\beta} = {\frac i2}\left(\gamma_\alpha \gamma_\beta - \gamma_\beta\gamma_\alpha\right)$ are the Lorentz algebra generators, where the flat Dirac matrices $\gamma^\alpha$ are defined in local Lorentz frames.

We denote the local spatial and time coordinates by $x^i = (t, x^a),$ $a,b,c =1, 2, 3$. An orthonormal coframe (tetrad) is needed to attach spinor spaces at every point of the space–time manifold. Then the dynamics of the Dirac particle can be investigated in an arbitrary Poincar\'e gauge field $(e^\alpha_i, \Gamma_i{}^{\alpha\beta})$, where the components of tetrads in the Schwinger gauge \cite{8} read
\begin{equation}\label{coframe}
e_i^{\,\widehat{0}} = V\,\delta^{\,0}_i,\qquad e_i^{\widehat{a}} =
W^{\widehat a}{}_b\left(\delta^b_i - cK^b\,\delta^{\,0}_i\right),\qquad a,b = 1,2,3.
\end{equation}
As was shown in Ref. \cite{8}, the Hermitian Hamiltonian of the fermion particle has the form
\begin{eqnarray}
{\cal H} &=& \beta mc^2V + q\Phi + {\frac c 2}\left(\pi_b
\,{\cal F}^b{}_a \alpha^a + \alpha^a{\cal F}^b{}_a\pi_b\right)\nonumber\\
&& +\,{\frac c2}\left(\bm{K}\!\cdot\bm{\pi} + \bm{\pi}\!\cdot\!\bm{K}\right) +
{\frac {\hbar c}4}\left(\bm{\Xi}\!\cdot\!\bm{\Sigma} - \Upsilon\gamma_5\right),
\label{Hamilton1}
\end{eqnarray}
where the  kinetic 3-momentum operator $\pi_a = -i\hbar\partial_a  - qA_a = p_a - qA_a$ accounts of the interaction with the electromagnetic field, and we denoted
\begin{equation}\label{AB}
{\cal F}^b{}_a = VW^b{}_{\widehat a},\qquad \Upsilon = V\varepsilon^{\widehat{a}\widehat{b}
\widehat{c}}\Gamma_{\widehat{a}\widehat{b}\widehat{c}},\qquad \Xi^a = {\frac Vc}\,
\varepsilon^{\widehat{a}\widehat{b}\widehat{c}}\left(\Gamma_{\widehat{0}\widehat{b}\widehat{c}} +
\Gamma_{\widehat{b}\widehat{c}\widehat{0}} + \Gamma_{\widehat{c}\widehat{0}\widehat{b}}\right).
\end{equation}
As usual,  $\alpha^a = \beta\gamma^a$ ($a,b,c,\dots = 1,2,3$) and the spin matrices
$\Sigma^1 = i\gamma^{\hat 2}\gamma^{\hat 3}, \Sigma^2 = i\gamma^{\hat 3}\gamma^{\hat 1},
\Sigma^3 = i\gamma^{\hat 1}\gamma^{\hat 2}$ and $\gamma_5=i\alpha^{\hat{1}}\alpha^{\hat{2}}
\alpha^{\hat{3}}$. Boldface notation is used for 3-vectors ${\bm K} = \{K^a\}, \,{\bm\pi}
= \{\pi_a\},\, {\bm\alpha} = \{\alpha^a\}, \,{\bm\Sigma} = \{\Sigma^a\}$.

Taking into account the decomposition of the connection (\ref{GGK}) into the Riemannian and post-Riemannian parts, we find that the Pauli-like equation with the Hermitian Hamiltonian (\ref{Hamilton1}) encompasses the spin-torsion coupling:
\begin{equation}\label{UX}
\Upsilon = \widetilde{\Upsilon} + Vc\check{T}^{\widehat{0}},\qquad
\Xi^{\widehat{a}} = \widetilde{\Xi}^{\widehat{a}} - V\check{T}^{\widehat{a}}.
\end{equation}
The tilde denotes the Riemannian quantities. The post-Riemannian contributions come from the components $\check{T}^\alpha = (\check{T}^{\widehat{0}}, \check{T}^{\widehat{a}})$ of the axial torsion vector (\ref{Taxial}). Accordingly, the spin-torsion coupling terms read explicitly
\begin{equation}
-\,{\frac {\hbar cV}4}\left(\bm{\Sigma}
\!\cdot\!\check{\bm{T}} + c\gamma_5\check{T}{}^{\hat 0}\right).\label{ST}
\end{equation}

The above general formalism can be applied to the study of fermion's dynamics in arbitrary external electromagnetic and gravitational (including the post-Riemannian one) fields.

Let us now specialize to the analysis of the possible physical effects of the spacetime torsion and the inertial forces on the non-relativistic particle in the rotating reference frame (such as the Earth), \cite{HN}:
\begin{equation}\label{VWni}
V = 1,\quad W^{\widehat a}{}_b = \delta^a_b,\quad K^a = -\,{\frac {(\bm{\omega}\times\bm{r})^a}{c}},
\qquad \Gamma_{\hat 0}{}^{\hat{a}\hat{b}} = -\,{\frac {\varepsilon^{abc}\omega_c}{c}},\quad
\Gamma_{\hat 0}{}^{\hat{a}\hat{0}} = 0.
\end{equation}
Substituting this configuration into the Hamiltonian (\ref{Hamilton1}) we derive
\begin{equation}\label{Hamlroton}
{\cal H} = \beta mc^2 + c\bm{\alpha}\cdot\bm{\pi} - \bm{\omega}\cdot(\bm{r}\times\bm{\pi}) 
- {\frac{\hbar}{2}}\bm{\omega}\cdot\bm{\Sigma} - {\frac{\hbar c}{4}}\left(
\check{T}^{\hat{0}} c\gamma_5 + \bm{\check{T}}\cdot\bm{\Sigma}\right).
\end{equation}
In order to reveal the physical contents of the Schr\"odinger equation, we need to go to the Foldy-Wouthuysen (FW) representation. Applying the methods developed in \cite{8}, we find the FW Hamiltonian
\begin{eqnarray}
H &=& \beta\epsilon + q\phi - \bm{\omega}\cdot(\bm{r}\times\bm{\pi}) 
- {\frac \hbar2}\bm{\omega}\cdot\bm{\Sigma} - {\frac{q\hbar c^2}{4}}\left\{{\frac{1}{\epsilon}},
\bm{B}\cdot\bm{\Pi}\right\}\nonumber\\
&& + \,{\frac{\hbar c^3}{8}}\left\{{\frac{\bm{\pi}\cdot\bm{\Pi}}{\epsilon}},\check{T}^{\hat{0}}
\right\} - {\frac{\hbar c}{8}}\left\{{\frac{mc^2}{\epsilon}},\bm{\check{T}}\cdot\bm{\Sigma}
\right\}\nonumber\\
&& -\,{\frac{\hbar c^3}{8}}\left[{\frac{\bm{\Sigma}\cdot\bm{\pi}}{\epsilon(\epsilon+mc^2)}}
\,\bm{\pi}\cdot\bm{\check{T}} + \bm{\check{T}}\cdot\bm{\pi}\,{\frac{\bm{\Sigma}\cdot\bm{\pi}}
{\epsilon(\epsilon+mc^2)}}\right]\nonumber\\
&& -\,{\frac{q\hbar c}{8}}\left\{{\frac{1}{\epsilon(\epsilon+mc^2)}},\bm{\Sigma}\cdot
(\bm{\mathfrak E}\times\bm{\pi} - \bm{\pi}\times\bm{\mathfrak E})\right\}.\label{HFW1}
\end{eqnarray}
Here $\bm{\Pi} = \beta\bm{\Sigma}$, $\{\ ,\ \}$ denotes anticommutators, $\epsilon=\sqrt{m^2c^4+c^2\bm{\pi}^2}$, and $\bm{\mathfrak E} = \bm{E} + \bm{B}\times(\bm{\omega}\times\bm{r})$ is the physical electric field as seen in the noninertial rotating reference frame. 

Under ordinary conditions we assume $|e\hbar B|\ll m^2c^2$, that is the magnetic field is much smaller than the critical field $|B|\ll B_c = m^2c^2/e\hbar$, and particle's velocity is much smaller than the speed of light, $|\pi|/m \ll c$. Then $\epsilon = mc^2 + \bm{\pi}^2/2m$, and in the semiclassical limit of (\ref{HFW1}) we finally obtain the Pauli-type equation $i\hbar\frac{\partial \psi} {\partial t} = H^{\rm nr}\psi$ with the non-relativistic Hamiltonian
\begin{equation}
H^{\rm nr} = {\frac {\bm{\pi}^2}{2m}} + q\phi - \bm{\omega}\cdot(\bm{r}\times\bm{\pi})
- {\frac \hbar2}\bm{\omega}\cdot\bm{\sigma} - {\frac{q\hbar}{2m}}\,\bm{B}\cdot\bm{\sigma}
+ {\frac{\hbar c}{8m}}\left\{\bm{\pi}\cdot\bm{\sigma}, \check{T}^{\hat{0}}\right\}
- {\frac{\hbar c}{4}}\,\bm{\check{T}}\cdot\bm{\sigma}.\label{Hnon}
\end{equation}
This result is consistent with an alternative analysis based on the method of exact FW transformations \cite{GOS}, see the relevant discussion in \cite{prl1,prl2}.

In the physically important situations, the torsion pseudovector is spacelike, and $|\check{\bm{T}}| \gg c\check{T}^{\hat{0}}$. Taking this into account, we now switch to the physically interesting case $\check{T}^{\hat{0}} = 0$. This is the true for the fermions (\ref{Ldir})-(\ref{eqin2}) minimally coupled to gravity.

\section{Quantum hydrodynamics for spin-torsion coupling}\label{Sec3}

In this section we derive the many-particle quantum hydrodynamics (MPQHD) equations from the many-particle Pauli-like equation for the system of charged particles with spin-$1/2$. The method of MPQHD allows to present the dynamics of a system of interacting quantum particles in terms of the functions defined in the three-dimensional physical space. This is important for the study of wave process, which take place in a three-dimensional physical space \cite{And,Truk}. In flat spacetime, the MPQHD formalism for many-particles fermion systems was previously developed in \cite{Kuz1,Kuz2,Kuz3}, whereas the case of the noninertial reference frames was considered in \cite{Truk2}. The methods of MPQHD can be used for the analysis of a wide variety of systems of many interacting particles. In particular, the finite temperature hydrodynamic model has been derived recently in \cite{Mos} for the spin-1 ultracold bosons.  In Ref. \cite{Dipole} the method was applied to the study of the polarization dynamics in a system of quantum particles with nontrivial electric dipole moments.

After applying the Foldy-Wouthuysen transformation for Dirac particle in combination with the method of many-particle quantum hydrodynamics, we arrive at the many-particle Pauli-like equation
\begin{equation}
i\hbar\frac{\partial \psi_s}{\partial t}=\hat{H}\psi_s,\label{Pauli}
\end{equation}
where the many-particle wave function of the system of $N$ spinning particles
\begin{equation}
\psi_s(R,t)=\psi_s(\bm{r}_1, \bm{r}_2,\dots,\bm{r}_N,t)
\end{equation}
is a spinor function in the $3N$-dimensional configuration space ($s$ is the spin index), and the many-particle Hamiltonian reads 
\begin{equation}\label{Hamiltonian1}
\hat{H} = \sum_{p=1}^N \left( {\frac{\hat{\pi}^2_p}{2m_p}}
+ {\frac \hbar 2}\,\bm{\sigma}\cdot\bm{\Omega}_{p} + \phi_p\right).
\end{equation}

Here we introduced
\begin{eqnarray}\label{Qp}
\bm{\Omega}_p &=& -\,\bm{\omega} - {\frac{q_p}{m_p}}\bm{B}_p - {\frac{c}{2}}\bm{\check{T}}{}_p,\\
\phi_p &=& q_p\,\phi(\bm{r}_p) - \frac{m_p}{2}[\bm{\omega}\times\bm{r}_p]^2,\label{phip}\\
\bm{\hat{\pi}}_p &=& -\,i\hbar\bm{\nabla}_p - q_p\bm{A}_p - m_p\,\bm{\omega}\times\bm{r}_p
\,,\label{covariant_derivative}
\end{eqnarray}
and $m_p$ and $q_p$ denote the mass and the charge of $p$-th particle, respectively. In particular, $q_p$ stands for the charge of electrons $q_e = -\,e$, or for the charge of ions $q_i = e$. The electromagnetic vector and scalar potentials $\bm{A}_p = \bm{A}(\bm{r}_p)$ and $\phi = \phi(\bm{r}_p)$ are taken at the positions $\bm{r}_p$ of the $p$-th particle, and the same applies to the external magnetic $\bm{B}_p = \bm{B}(\bm{r}_p)$ and the torsion $\bm{\check{T}}{}_p = \bm{\check{T}}(\bm{r}_p)$ fields. The last terms in (\ref{phip}) and (\ref{covariant_derivative}) manifest the inertial contributions in the rotating reference frame with the angular velocity $\bm{\omega}$.

As compared to the standard case of a system in an external electromagnetic field, the many-particle Hamiltonian (\ref{Hamiltonian1}) includes the torsion effects, encoded in the second term $\sim \bm{\sigma}\cdot\bm{\check{T}}{}_p$, that has the same form as the Zeeman energy in the magnetic field. In addition, this Hamiltonian includes Mashhoon's spin-rotation contribution $\sim \bm{\sigma}\cdot\bm{\omega}$, see \cite{Mashhoon1,Mashhoon2,Mashhoon3}. 

The state of the system is characterized by the concentration of particles in the neighborhood of a point $\bm{r}$ in the physical space as
\begin{equation} \label{n}
n(\bm{r},t)=\int dR \sum_{p=1}^N\delta(\bm{r}-\bm{r}_p)\psi^*_s(R,t)\psi_s(R,t)
=\langle\psi^{\dagger}\hat{n}\psi\rangle.
\end{equation}
Here the integration measure reads $dR = \prod_p\,d^3r_p$. The function $n(\bm{r},t)$ is thus determined as the quantum average of the concentration operator $\hat{n}=\sum_p\delta(\bm{r}-\bm{r}_p)$ in the coordinate representation. The spin density vector of fermions is determined in a similar way
\begin{equation}\label{spin}
\bm{S}(\bm{r},t)=\int dR \sum_{p=1}^N\delta(\bm{r}-\bm{r}_p)\psi^*_s(R,t)
(\bm{\hat{s}}_p)_{ss'}\psi_{s'}(R,t) = \langle\psi^{\dagger}\bm{\mathfrak{s}}\psi\rangle,
\end{equation}
as the quantum average of the spin operator $\bm{\mathfrak{s}} = \sum_p\delta(\bm{r}-\bm{r}_p)\bm{\hat{s}}_p$, with $\bm{\hat{s}}_p = {\frac{\hbar}{2}}\bm{\sigma}_p$.

The continuity equation for the concentration of the particles $n(\bm{r},t)$ can be derived by taking the time derivative of the definition (\ref{n}) and making use of the many-particle Pauli-like equation (\ref{Pauli}):
\begin{equation}\label{nn}
  \partial_t n(\bm{r},t) + \bm{\nabla}\cdot\bm{J}(\bm{r},t) = 0,
\end{equation}
where the current density is defined as the microscopic average $\bm{J}(\bm{r},t) = {\frac 12}\langle\psi^{\dagger}\bm{\mathfrak J}\psi + {\rm c.c.}\rangle$ of the operator
\begin{equation} \label{current}
\bm{\mathfrak{J}} = \sum_p\delta(\bm{r}-\bm{r}_p)\,{\frac{\bm{\hat{\pi}}_p}{m_p}}.
\end{equation}
Here the generalized momentum operator is defined by (\ref{covariant_derivative}).

\subsection{Spin density evolution}

In a similar way, the dynamical equation for the spin density can be obtained by differentiating the definition (\ref{spin}) with respect to time and making use of the many-particle Pauli-like equation (\ref{Pauli}):
\begin{equation}\label{spin_equation}
\partial_tS^a(\bm{r},t) + \partial_b\Lambda^{ba}(\bm{r},t) = \varepsilon^{abc}\Omega^bS^c(\bm{r},t).
\end{equation}
Here the spin precession angular velocity is defined as
\begin{equation}\label{Omega1}
\bm{\Omega} = -\,\bm{\omega} - {\frac{q}{m}}\bm{B} - {\frac{c}{2}}\bm{\check{T}},
\end{equation}
cf. (\ref{Qp}), whereas the spin current density tensor is introduced as a microscopic average $\Lambda^{ba}(\bm{r},t) = {\frac 12}\langle\psi^{\dagger}{\mathfrak L}^{ba}\psi + {\rm c.c.}\rangle$ of the operator 
\begin{equation}\label{spin_current}
\mathfrak{L}^{ba} = \sum_p\delta(\bm{r}-\bm{r}_p)\,{\frac{\hat{\pi}^a_p\hat{s}^b_p}{m_p}}.
\end{equation}

\subsection{Equation of motion}

Along the same lines, the derivation of the equation of motion of a hydrodynamic system in an external electromagnetic and torsion fields is based on differentiating the expression for the current density $\bm{J}(\bm{r},t)$ with respect to time and using the Pauli-like equation (\ref{Pauli}) with the Hamiltonian (\ref{Hamiltonian1}). The result reads
\begin{equation}\label{current_equation}
m\partial_tJ^a(\bm{r},t) + \partial_b\Pi^{ab}(\bm{r},t) = qnE^a(\bm{r},t)
+ q\varepsilon^{abc}J^b(\bm{r},t)B^c(\bm{r},t) - S^b(\bm{r},t)\partial^a\Omega^b + F^a_{\rm iner},
\end{equation}
where the momentum flux tensor appears in fluid dynamics as a quantum average $\Pi^{ab}(\bm{r},t) = {\frac 12}\langle\psi^{\dagger}{\mathfrak P}^{ab}\psi + {\rm c.c.}\rangle$ of the operator 
\begin{equation}\label{flux}
\mathfrak{P}^{ab} = \sum_p\delta(\bm{r}-\bm{r}_p)\,{\frac{\hat{\pi}^{(a}_p\hat{\pi}^{b)}_p}{m_p}}.
\end{equation}
The equation of motion (\ref{current_equation}) describes the influence of the external electromagnetic and torsion fields on the fermion matter in terms of the Lorentz and the Stern-Gerlach forces.

As a next step, one can move from the microscopic representation of the particle current density and the spin current density to their corresponding macroscopic variables by making use of an explicit representation of the spinor wave function. Such an explicit representation of the wave function is known as the Madelung decomposition.

\section{Madelung decomposition}\label{Sec4} 

The microscopic many-particle wave function or the Madelung decomposition \cite{Madelung} of the $N$-particle wave function can be represented in terms of the amplitude $a(R,t)$, the phase $\xi(R,t)$ and the local spinor $\mathcal{Z}(R,\bm{r},t)$, defined in the local rest frame and normalized so that $\mathcal{Z}^{\dagger}\mathcal{Z}=1$:
\begin{equation} \label{wave3}
\psi(R,t) = a(R,t)\,\varphi(R,\bm{r},t),\qquad \varphi(R,\bm{r}.t) =
e^{{\frac{i}{\hbar}}\,\xi(R,t)}\,\mathcal{Z}(R,\bm{r},t).
\end{equation}
Applying the  decomposition (\ref{wave3}) to the $p$-th particle, we can introduce a microscopic velocity and a microscopic spin as $\bm{v}_p := {\frac 1{m_p}}\varphi^\dagger\,\bm{\hat{\pi}}_p\,\varphi$ and $\bm{s}_p := \varphi^\dagger\,\bm{\hat{s}}_p\,\varphi = {\frac {\hbar}{2}}\,\varphi^\dagger\,\bm{\sigma}_p\,\varphi$, respectively. Explicitly, we then find
\begin{eqnarray}\label{micv}
\bm{v}_p(R,\bm{r},t) &=& {\frac{1}{m_p}}\left(\bm{\nabla}_p\,\xi - q\bm{A}_p
-i\hbar\mathcal{Z}^{\dagger}\bm{\nabla}_p\,\mathcal{Z} - m_p\,\bm{\omega}\times\bm{r}_p\right),\\
\bm{s}_p(R,\bm{r},t) &=& {\frac{\hbar}{2}}\,\mathcal{Z}^{\dagger}\bm{\sigma}_p\mathcal{Z}.\label{mics}
\end{eqnarray}
The velocity field of the $p$-th particle can be decomposed $\bm{v}_p(R,\bm{r},t) = \bm{v}(\bm{r},t) + \bm{\eta}_p(R,\bm{r},t)$ into a sum of the macroscopic average $\bm{v}(\bm{r},t)$ and the thermal fluctuations part $\bm{\eta}_p(R,\bm{r},t)$ of the velocity. In a similar way, the spin of the $p$-th particle can be represented as the sum $\bm{s}_p(R,\bm{r},t) = \bm{s}(\bm{r},t) + \bm{\tau}_p(R,\bm{r},t)$ of the macroscopic average $\bm{s}(\bm{r},t)$ and the thermal fluctuations part $\bm{\tau}_p(R,\bm{r},t)$ of the spin. By definition, the averages of the fluctuations vanish, $\langle a^2\bm{\eta}_p\rangle=0$ and $\langle a^2\bm{\tau}_p\rangle=0$. We assume that the particle system is closed and not placed in a thermostat. Recalling that the temperature is the average kinetic energy of the chaotic motion of the particles of the system, we consider deviations of the velocity and spin of quantum particles from the local average values, which correspond to the ordered motion of the particles.

Combining Eqs. (\ref{n}), (\ref{current}), and (\ref{spin}) with (\ref{wave3})-(\ref{mics}), we can derive the macroscopic concentration, the macroscopic current density and the macroscopic spin density from the corresponding microscopic variables:
\begin{align}\label{concentration1}
n(\bm{r},t) &= \int dR \sum_{p=1}^N\delta(\bm{r}-\bm{r}_p)\,a^2(R,t),\\
\bm{J}(\bm{r},t) &= \int dR \sum_{p=1}^N\delta(\bm{r}-\bm{r}_p)\,a^2(R,t)\,\bm{v}_p(R,\bm{r},t)
= n(\bm{r},t)\,\bm{v}(\bm{r},t),\label{current1}\\
\bm{S}(\bm{r},t) &= \int dR \sum_{p=1}^N\delta(\bm{r}-\bm{r}_p)\,a^2(R,t)\,\bm{s}_p(R,\bm{r},t)
= n(\bm{r},t)\,\bm{s}(\bm{r},t).\label{spin1}
\end{align}

After the Madelung decomposition procedure for the basic physical variables in the microscopic representation, the spin current density (\ref{spin_current}) and the momentum flux (\ref{flux}) can be recast in terms of the fluid variables into
\begin{eqnarray}
\Lambda^{ba}(\bm{r},t) &=& \int dR \sum_{p=1}^N\delta(\bm{r}-\bm{r}_p)\left(
a^2s^a_pv^b_p-\frac{a^2}{m_p}\varepsilon^{acd}s^c_p\partial^b_ps^d_p\right),\\
\Pi^{ab}(\bm{r},t) &=& \int dR \sum_{p=1}^N\delta(\bm{r}-\bm{r}_p)\left({\frac{\hbar^2}{2m_p}}
\,(\partial^a_pa\partial^b_pa - a\partial^a_p\partial^b_pa) \right.\nonumber\\ 
&&\left. + \,m_p a^2 v^a_p v^b_p + {\frac{a^2}{m_p}}\,\partial^a_ps^c_p\partial^b_ps^c_p\right),
\end{eqnarray}
respectively. We are now ready to write down the complete set of dynamical equations for the quantum system of spinning particles explicitly in terms of the fluid variables. This set encompasses the continuity equation
\begin{equation}\label{continuity_equation}
\partial_t n + \bm{\nabla}\cdot(n\,\bm{v}) = 0,
\end{equation}
the momentum balance equation
\begin{eqnarray}
(\partial_t + v^b\partial_b)\,v^a &=& {\frac{q}{m}}E^a + {\frac{q}{m}}(\bm{v}\times\bm{B})^a
- {\frac{1}{n}}\partial_bp^{ab} + {\frac{\hbar^2}{2m^2}}\,\partial^a\biggl(\frac{\Delta
\sqrt{n}}{\sqrt{n}}\biggr)\nonumber\\
&& + \,{\frac{1}{2m^2}}\partial^a(\partial^b\bm{s}\cdot\partial^b\bm{s}) -
{\frac{1}{m}}\bm{s}\cdot\partial^a\bm{\hat{\Omega}} 
+ f^a_{\rm iner} - {\frac{1}{m}}\,Q^a_{\rm therm},\label{momentum_equation}
\end{eqnarray}
whereas the spin evolution equation (\ref{spin_equation}) reads
\begin{equation}\label{spin_equation2}
(\partial_t + v^b\partial_b)\,\bm{s} = \bm{\hat{\Omega}}\times\bm{s} -\bm{\Theta}_{\rm therm},
\end{equation}
where the spin precession angular velocity is modified, cf. (\ref{Omega1}),
\begin{equation}\label{Omega2}
\bm{\hat{\Omega}} = -\,\bm{\omega} - {\frac{q}{m}}\bm{B} - {\frac{c}{2}}\bm{\check{T}}
- {\frac {1}{mn}}\,\partial_b(n\partial^b\bm{s}). 
\end{equation}
This modification arises from the interaction of spin with the surrounding spin-texture of the fluid, and one can formally interpret this in terms of an effective magnetic field defined as a sum of an external magnetic field and the emergent field 
\begin{equation}\label{magnetic_field}
\bm{\hat{B}} = \bm{B} + {\frac{1}{qn}}\,\partial_b(n\partial^b\bm{s}). 
\end{equation}
In fact, one can also view the first term on the right-hand side of (\ref{Omega2}), which is due to Mashhoon's spin-rotation coupling term and describes the Barnett effect, and the third term on the right-hand side of (\ref{Omega2}), representing the spin-torsion coupling, as the two additional contributions to the ``effective'' magnetic field
\begin{equation}
\bm{B}_{\omega} = {\frac{m}{q}}\,\bm{\omega},\qquad \bm{B}_{T} = {\frac{mc}{2q}}\,\bm{\check{T}}.
\end{equation}

The dynamical equation (\ref{spin_equation2}) describes the precession of spin under the action of the torque produced by the external magnetic field and the emergent fields, leading to the Zeeman type effect. The additional torque in (\ref{spin_equation2}) arises from thermal-spin interactions
\begin{equation}
\bm{\Theta}_{\rm therm} = \partial_b \int dR \sum_{p=1}^N\delta(\bm{r}-\bm{r}_p)\left(
a^2\eta^b_p \bm{s}_p - {\frac{a^2}{m_p}}\,\bm{s}_p\times\partial^b_p\bm{\tau}_p\right),
\end{equation}
that is also responsible for the last force term in the momentum balance equation (\ref{momentum_equation})
\begin{eqnarray}
Q^a_{\rm therm} &=& {\frac 1n}\,\partial_b \int dR \sum_{p=1}^N\delta(\bm{r}-\bm{r}_p)\,\frac{a^2}{m_p}
\left(\partial_p^b\bm{\tau}_p\cdot\partial_p^a\bm{s}_p +\partial_p^b \bm{s}_p\cdot\partial^a_p\bm{\tau}_p
-\partial_p^a\bm{\tau}_p\cdot\partial_p^b\bm{\tau}_p\right)\nonumber\\
&& - \,\partial^a\left\{\frac{1}{n}\,\partial_b\biggl[n\,\partial^b \Bigl(\frac{1}{n} \int dR
\sum_{p=1}^N\delta(\bm{r}-\bm{r}_p)\,\frac{a^2}{2m_p}\,\bm{\tau}_p\cdot\bm{\tau}_p\Bigr)\biggr]\right\}.
\end{eqnarray}

Analysing the structure of the equation of motion (\ref{momentum_equation}), we identify the first two terms on the right hand side with the Lorenz force determined by the external electric and magnetic fields $\bm{E}$ and $\bm{B}$, while the third term is the divergence of the kinetic pressure tensor
\begin{equation}
p^{ab}(\bm{r},t) = \int dR \sum_{p=1}^N\delta(\bm{r}-\bm{r}_p)\,a^2\,m_p\,\eta^a_p\,\eta^b_p.
\end{equation}
The fifth term on the right hand side of the equation (\ref{momentum_equation}) represents the effect of spin-spin interactions inside the fluid, the interaction of the spin with the spin background texture. The sixth term describes the Stern-Gerlach force that characterizes the influence of the non-uniform effective magnetic and the torsion field. In the non-inertial frame, an additional contribution encompasses the Coriolis force, the centrifugal force and Euler force field
\begin{equation}
\bm{f}_{\rm iner} = -\,2\,\bm{\omega}\times\bm{v}-\bm{\omega}\times(\bm{\omega}\times\bm{\mathfrak R})
- {\frac{\partial\bm{\omega}}{\partial t}}\times\bm{\mathfrak R},
\end{equation}
where the vector of center of mass is defined as
\begin{equation}
  \bm{\mathfrak R}(\bm{r},t) = {\frac 1n}\int dR \sum_{p=1}^N\delta(\bm{r}-\bm{r}_p)
  \psi^*_s(R,t)\bm{r}_p\psi_s(R,t).
\end{equation}
Our derivations are consistent with the earlier analysis \cite{Truk2}. 

\section{Experimental manifestations of spin-torsion coupling}\label{Sec5}

Experimental search for the nontrivial torsion effects is naturally embedded into the broader framework of the studies of the spin-dependent long-range forces \cite{93,94,95,96,10,11,12,13,GPS,14,Mohanty,Belyaev,VAK,VAK2,Heckel,Kimball,Ivanov1,Ivanov2,Fabbri,15}. By making use of the corresponding experimental techniques, it is possible to find strong limits on the values of the gauge gravity spin-torsion coupling constants and on the torsion field itself. A good example of an efficient approach in this respect gives an observation of the nuclear spin precession in gaseous spin polarized $^3$He or $^{129}$Xe samples with the help of a highly sensitive low-field magnetometer \cite{16,17,18} detecting a sidereal variation of the relative spin precession frequency in a new type of $^{3}$He/$^{129}$Xe clock comparison test. In a similar experiment \cite{Venema}, the ratio of nuclear spin-precession frequencies of $^{199}$Hg and $^{201}$Hg atoms was measured in the magnetic field and the Earth's gravitational field. Based on the corresponding experimental data from \cite{Venema}, the analysis of dynamics of the minimally coupled Dirac fermion \cite{8} in external electromagnetic and gravitational fields revealed the strong bounds on the possible background space-time torsion:
\begin{equation}
{\frac c2}\,|\bm{\check{T}}|\cdot |\cos\theta| < 6.45\times 10^{-6}s^{-1},
\end{equation}
where $\theta$ is the angle between the magnetic $\bm{B}$ and torsion $\bm{\check{T}}$ fields. On the other hand, by making use of the experimental data from \cite{16}, one finds the restriction 
\begin{equation}
{\frac c2}\,|\bm{\check{T}}|\cdot |\cos\theta| < 3.59\times 10^{-7}s^{-1}.
\end{equation}
These results are consistent with the alternative empirical estimates for the torsion limits \cite{14,Mohanty,Belyaev,VAK,VAK2,Heckel,Kimball,Ivanov1,Ivanov2,Fabbri,15}. As another powerful tool one can mention the use of the quantum interferometry to probe the spacetime structure, including the search for possible post-Riemannian deviations, focusing on the detection of the phase shift and polarization rotation effects for the neutron and atom beams \cite{19,20,Kiefer,Lamm,neutron}.

As an application of the quantum hydrodynamics formalism, let us investigate a simple model of a continuous medium of particles with spin and consider the dynamics of the spin waves in such a particle system. Neglecting the spin-thermal coupling in the spin dynamical equation (\ref{spin_equation2}) and assuming the small perturbations of the spin $\bm{s} = \bm{s}_0 + \delta \bm{s}$ around an undisturbed value $|\bm{s}_0| = \hbar/2$, we find, in the first order,
\begin{equation}
\partial_t\delta \bm{s}=\bm{\hat{\Omega}}^{(0)}\times\delta\bm{s}+\bm{\hat{\Omega}}^{(1)}\times\bm{s}_0.
\end{equation}
Here the equilibrium values of the external background fields are encoded in $\bm{\hat{\Omega}}^{(0)} = -\,\bm{\omega}_0 - {\frac{q}{m}}\bm{B}_0 - {\frac{c}{2}}\bm{\check{T}}$, where $\bm{B}_0$, $\bm{\omega}_0$ and $\bm{\check{T}}$ are the external uniform magnetic field, the Earth's angular velocity and the background torsion, respectively, and the small disturbance reads
\begin{equation}
\bm{\hat{\Omega}}^{(1)} = -\,{\frac{\Delta\delta\bm{s}}{m}},
\end{equation}
Assuming that the perturbations of the spin vary as $\delta\bm{s} \sim \exp(- i\omega_s t + i\bm{k}\cdot\bm{r})$, we then derive the dispersion law relating the wave frequency $\omega_s$ and the wave vector $\bm{k}$ for the spin waves excited in the external magnetic and torsion fields:
\begin{equation}\label{w}
\omega_s^2 = \Omega_c^2 + {\frac{c^2}{4}}\check{T}^2 + \omega_0^2 + 2\Omega_c\omega_0\cos\theta_1
+ c\,\Omega_c\check{T}\cos\theta_2 + c\,\omega_0\check{T}\cos(\theta_2-\theta_1) .
\end{equation}
Here $\Omega_c = {\frac {qB_0}{m}} + {\frac{\hbar k^2}{2m}}$, whereas $\theta_1$ and $\theta_2$ are, respectively, the angle between the external magnetic field and Earth's angular velocity, and the angle between $\bm{B}_0$ and the background torsion $\bm{\check{T}}$. Equation (\ref{w}) is a generalization for the dispersion relation of spin waves found in Ref. \cite{MISRA} and it takes into account the contribution of the spin part of the quantum Bohm potential as an additional spin torque due to the self-action inside the system of particles, which leads to the propagation of spin waves. The square of the frequency $\omega_s^2$ encompasses a contribution proportional to the square of the modulus of the wave vector $\sim k^2$. As we can see from the dispersion relation (\ref{w}) the torsion effect is maximal when the pseudovector field $\bm{\check{T}}$ is aligned along the external magnetic field.

\section{Discussion and conclusion}\label{Sec6}

In this paper we for the first time developed the quantum hydrodynamics for the many-particle system of massive Dirac fermion spin-1/2 particles interacting with external electromagnetic, metric gravitational/inertial and torsion fields. This essentially extends the single-particle quantum hydrodynamical approach which was developed for the flat spacetime, see \cite{Takabayasi1,Takabayasi2,Recami,Fabbri2,Hestenes,Holland} and the references therein. Taking as the basis of the earlier general formalism \cite{8}, the consistent hydrodynamical formulation was constructed for the many-particle quantum system of fermions, and the explicit relations between the microscopic and macroscopic fluid variables were derived with help of the Madelung decomposition approach. In the present study, we have focused on the physically important situation with $\check{T}^{\hat{0}} = 0$, a more exotic case with $\check{T}^{\hat{0}} \neq 0$ (which may be realized in cosmology, for example, see the earlier work \cite{9}, or in the more general models) will be considered elsewhere.

The resulting system of hydrodynamical equations consists of the continuity equation (\ref{continuity_equation}), the momentum balance equation (\ref{momentum_equation}) and the spin dynamics equation (\ref{spin_equation2}). The momentum balance equation includes the contributions in the form of the quantum Bohm potential and a new spin part of the Bohm quantum potential which are proportional to the square of Planck's constant. In addition, the dynamical equations take into account the thermal effects resulting from the fluctuations of the spin and velocity near their average values. As an application of the formalism, we evaluated the possible effects of the spacetime torsion and the spin part of quantum Bohm potential on the dispersion characteristics of the spin waves (\ref{w}) excited in the many-particle fermion system. The developed hydrodynamical model can be used in the future studies of various types of transport phenomena in spinning matter with an account of external electromagnetic and gravitational fields. 

\section*{Acknowledgments}

The work of MIT was supported by the Russian Science Foundation under the grant 19-72-00017, and the research of YNO was supported in part by the Russian Foundation for Basic Research (Grant No. 18-02-40056-mega).


\end{document}